\documentclass[twocolumn,a4paper]{article}
\usepackage{txfonts}
\usepackage{amssymb}
\usepackage{graphicx}
\usepackage{dsfont}
\usepackage{stmaryrd}

\usepackage[T1]{fontenc}
\usepackage[utf8]{inputenc}

\usepackage{savesym}
\savesymbol{iint}
\savesymbol{iiint}
\savesymbol{iiiint}
\savesymbol{idotsint}
\savesymbol{openbox}
\usepackage{amsmath}
\usepackage{amscd}

\usepackage{amsthm}

\newtheorem{definition}{Definition}
\newtheorem{proposition}{Proposition}
\newtheorem{theorem}{Theorem}

\usepackage{subfigure}

\begin{document}

\title{Randomness and disorder of chaotic iterations. Applications in information security field}
\author{ Xiaole Fang ${}^\S$, Christophe Guyeux${}^\dag$, Qianxue Wang${}^\ddag$, and Jacques M. Bahi${}^\dag$}

\maketitle
\abstract
Design and cryptanalysis of chaotic encryption schemes are major concerns to provide secured information systems. Pursuing our previous research works, some well-defined discrete chaotic iterations that satisfy the reputed Devaney's definition of chaos have been proposed. In this article, we summarize these contributions and propose applications in the fields of pseudorandom number generation, hash functions, and symmetric cryptography. For all these applications, the proofs of chaotic properties are outlined.
\endabstract

\section{Introduction}
Applying chaotic systems to construct cryptosystems has been extensively investigated since 1990s, and this field of research has attracted more and more attention in the near decades. Some researchers have pointed out that there exists tight relationship between chaos and randomness, thus it is a natural idea to use chaos to enrich the design of cryptographic applications. However, almost all current researches of chaotic systems consider real domain. Since all operations (iterations) are on the real numbers, Real Domain Chaotic Systems (RDCSs) realized in a computer or a digital device will inevitably lead to finite precision effects, and may result in consequent dynamic degradation, such as short cycle-length, non-ideal distribution and correlation, low linear complexity, and so on. Chaotic iterations (CIs), for its part, refers to chaotic systems defined on integer domain. They have been deeply studied in our previous collaborative works, in order to solve degradation of chaotic dynamic properties by finite precision effects on traditional RDCSs. In this research work, we intend to deepen the theoretical and practical knowledge already obtained on CIs. More general theoretical designs and applications for security will be done to further investigate and learn more about the CIs framework.

The remainder of this research work is organized as follows. The basic recalls
of CIs are given in Section~\ref{Basic recalls}, while results of Devaney's chaos are provided in Section~\ref{Chaos results about chaotic iterations}. Sections \ref{Application to pseudorandom number generation}, \ref{Application to hash functions},
and~\ref{Application to symmetric cryptography} show applications to  pseudorandom number generation, hash functions, and symmetric cryptography respectively. This article ends by a conclusion section in which the article is summarized.

\section{Basic recalls}
\label{Basic recalls}
\subsection{Devaney's theory of chaos}

\label{subsec:Devaney}
In the remainder of this article, $S^{n}$ denotes the $n^{th}$ term of a sequence $S$ while $\mathcal{X}^\mathds{N}$ is the set of all sequences whose elements belong to $\mathcal{X}$. $V_{i}$
stands for the $i^{th}$ component of a vector $V$. $f^{k}=f\circ ...\circ f$
is for the $k^{th}$ composition of a function $f$. $\mathds{N}$ is the set of natural (non-negative) numbers, while $\mathds{N}^*$ stands for the positive integers $1, 2, 3, \hdots$ Finally, the following
notation is used: $\llbracket1;N\rrbracket=\{1,2,\hdots,N\}$.

Consider a topological space $(\mathcal{X},\tau)$ and a continuous function $f :
\mathcal{X} \rightarrow \mathcal{X}$ on $(\mathcal{X},\tau)$.

\begin{definition}
The function $f$ is \emph{topologically transitive} if, for any pair of open sets
$U,V \subset \mathcal{X}$, there exists an integer $k>0$ such that $f^k(U) \cap V \neq
\varnothing$.
\end{definition}

\begin{definition}
An element $x$ is a \emph{periodic point} for $f$ of period $n\in \mathds{N}$, $n>1$,
if $f^{n}(x)=x$. \linebreak 
%
$f$ is  \emph{regular} on $(\mathcal{X}, \tau)$ if the set of periodic
points for $f$ is dense in $\mathcal{X}$: for any point $x$ in $\mathcal{X}$,
any neighborhood of $x$ contains at least one periodic point.
\end{definition}

\begin{definition}[Devaney's formulation of chaos~\cite{Devaney}]
The function $f$ is  \emph{chaotic} on $(\mathcal{X},\tau)$ if $f$ is regular and
topologically transitive.
\end{definition}

Banks \emph{et al.} have proven in~\cite{Banks92} that, when the topological space is a metric one $(\mathcal{X},d)$, chaos implies sensitivity, defined below:
\begin{definition}
\label{sensitivity} The function $f$ has \emph{sensitive dependence on initial conditions}
if there exists $\delta >0$ such that, for any $x\in \mathcal{X}$ and any
neighborhood $V$ of $x$, there exist $y\in V$ and $n > 0$ such that
$d\left(f^{n}(x), f^{n}(y)\right) >\delta .$\\
$\delta$ is called the \emph{constant of sensitivity} of $f$.
\end{definition}



\subsection{Chaotic Iterations}

Define by $\mathcal{S}_X$ the set of sequences whose elements belong in $X \subset \mathds{N}, X \neq \varnothing$,
that is, $\mathcal{S}_X = X^\mathds{N}$.
\begin{definition}
\label{Def:chaotic iterations}
The      set       $\mathds{B}$      denoting      $\{0,1\}$,      let $\mathsf{N} \in
\mathds{N}^*$,
$f:\mathds{B}^{\mathsf{N}}\longrightarrow  \mathds{B}^{\mathsf{N}}$ be
a  function,  and  $S\in  \mathcal{S}_{\llbracket 1, \mathsf{N} \rrbracket}$  be  a sequence of integers between 1 and $\mathsf{N}$.  The  so-called
\emph{chaotic      iterations}     are     defined      by     $x^0\in
\mathds{B}^{\mathsf{N}}$ and
\begin{equation}
\forall    n\in     \mathds{N}^{\ast     },    \forall     i\in
\llbracket1;\mathsf{N}\rrbracket ,x_i^n=\left\{
\begin{array}{ll}
  x_i^{n-1} &  \text{ if  }S^n\neq i \\
  \left(f(x^{n-1})\right)_{S^n} & \text{ if }S^n=i.
\end{array}\right.
\end{equation}
\end{definition}
In other words, at the $n^{th}$ iteration, only the $S^{n}-$th component of the vector $x^n$ is
updated.  Note  that in  a more
general  formulation, each $S^n$  can   be  a  subset  of  $\{1, 2, \hdots, \mathsf{N}\}$. 
Let us  remark that
the term  ``chaotic'', in  the name of  these iterations,  has \emph{a
priori} no link with the mathematical theory of chaos, recalled before.

\section{Chaos results about chaotic iterations}
\label{Chaos results about chaotic iterations}
We now recall how to define a suitable metric space where chaotic iterations
are continuous. For further explanations, see, \emph{e.g.}, \cite{guyeux10}.
Let $\delta $ be the \emph{discrete Boolean metric}, $\delta
(x,y)=0\Leftrightarrow x=y.$ Given a function $f$, define the function
$F_{f}:  \llbracket1;\mathsf{N}\rrbracket\times \mathds{B}^{\mathsf{N}} 
\longrightarrow  \mathds{B}^{\mathsf{N}}$ by:
\begin{equation*} 
\begin{array}{lrll}
& (k,E) & \longmapsto & \left( E_{j}.\delta (k,j)+ f(E)_{k}.\overline{\delta
(k,j)}\right) _{j\in \llbracket1;\mathsf{N}\rrbracket}%
\end{array}%
\end{equation*}%
\noindent where + and . are the Boolean addition and product operations.
Consider the phase space:
$\mathcal{X} = \llbracket 1 ; \mathsf{N} \rrbracket^\mathds{N} \times
\mathds{B}^\mathsf{N}$, and the map defined on $\mathcal{X}$ by:
\begin{equation}
G_f\left(S,E\right) = \left(\sigma(S), F_f(i(S),E)\right), \label{Gf}
\end{equation}
\noindent where $\sigma$ is the \emph{shift} function defined by $\sigma :
(S^{n})_{n\in \mathds{N}}\in \llbracket 1, \mathsf{N} \rrbracket^\mathds{N}\longrightarrow (S^{n+1})_{n\in
\mathds{N}}\in \llbracket 1, \mathsf{N} \rrbracket^\mathds{N}$ and $i$ is the \emph{initial function} 
$i:(S^{n})_{n\in \mathds{N}} \in \llbracket 1, \mathsf{N} \rrbracket^\mathds{N}\longrightarrow S^{0}\in \llbracket
1;\mathsf{N}\rrbracket$. Then the chaotic iterations proposed in
Definition \ref{Def:chaotic iterations} can be described by the following discrete dynamical system, whose topological chaos can now be studied:
\begin{equation}
\left\{
\begin{array}{l}
X^0 \in \mathcal{X} \\
X^{k+1}=G_{f}(X^k).%
\end{array}%
\right.
\end{equation}%
To do so, a relevant distance between two points $X = (S,E), Y =
(\check{S},\check{E})\in
\mathcal{X}$ has been introduced in \cite{guyeux10} as follows:
$d(X,Y)=d_{e}(E,\check{E})+d_{s}(S,\check{S})$,
where
\begin{equation}
\left\{
\begin{array}{lll}
\displaystyle{d_{e}(E,\check{E})} & = & \displaystyle{\sum_{k=1}^{\mathsf{N}%
}\delta (E_{k},\check{E}_{k})}, \\
\displaystyle{d_{s}(S,\check{S})} & = & \displaystyle{\dfrac{9}{\mathsf{N}}%
\sum_{k=1}^{\infty }\dfrac{|S^k-\check{S}^k|}{10^{k}}}.%
\end{array}%
\right.
\end{equation}


It has been established in \cite{guyeux10} that,
\begin{proposition}
$G_{f}$ is continuous in
the metric space $(\mathcal{X},d)$.
\end{proposition}

The chaotic property of $G_f$ has been firstly established for the vectorial
Boolean negation $f_0(x_1,\hdots, x_\mathsf{N}) =  (\overline{x_1},\hdots, \overline{x_\mathsf{N}})$ \cite{guyeux10}. To obtain a characterization, we have secondly
introduced the notion of asynchronous iteration graph recalled thereafter~\cite{bcgr11:ip}.
Let $f$ be a map from $\mathds{B}^\mathsf{N}$ to itself. The
{\emph{asynchronous iteration graph}} associated with $f$ is the
directed graph $\Gamma(f)$ defined by: the set of vertices is
$\mathds{B}^\mathsf{N}$; for all $x\in\mathds{B}^\mathsf{N}$ and 
$i\in \llbracket1;\mathsf{N}\rrbracket$,
the graph $\Gamma(f)$ contains an arc from $x$ to $F_f(i,x)$. 
We have then proven in~\cite{bcgr11:ip} that,
\begin{theorem}
\label{Th:Caractérisation   des   IC   chaotiques}  
Let $f:\mathds{B}^\mathsf{N}\to\mathds{B}^\mathsf{N}$. $G_f$ is chaotic  (according to  Devaney) 
if and only if $\Gamma(f)$ is strongly connected.
\end{theorem}
Finally, we have established in \cite{bcgr11:ip} that,
\begin{theorem}
  Let $f: \mathds{B}^{n} \rightarrow \mathds{B}^{n}$, $\Gamma(f)$ its
  iteration graph, $\check{M}$ its adjacency
  matrix and $M$
  a $n\times n$ matrix defined by 
  $
  M_{ij} = \frac{1}{n}\check{M}_{ij}$ 
  if $i \neq j$ and  
  $M_{ii} = 1 - \frac{1}{n} \sum\limits_{j=1, j\neq i}^n \check{M}_{ij}$ otherwise.
  
  If $\Gamma(f)$ is strongly connected, then 
  the output of the chaotic iterations 
  follows 
  a law that tends to the uniform distribution 
  if and only if $M$ is a double stochastic matrix.
\end{theorem} 
These results of topological chaos and uniform distribution have led us to study the possibility of building a
pseudorandom number generator (PRNG) based on chaotic iterations. 
As $G_f$, defined on the domain   $\llbracket 1 ;  \mathsf{N} \rrbracket^{\mathds{N}} 
\times \mathds{B}^\mathsf{N}$, is built from Boolean networks $f : \mathds{B}^\mathsf{N}
\rightarrow \mathds{B}^\mathsf{N}$, we can preserve the theoretical properties on $G_f$
during implementations (due to the discrete nature of $f$). 

\section{Application to pseudorandom number generation}
\label{Application to pseudorandom number generation}

Let $\mathsf{N} \in \mathds{N}^\ast$, $f:\mathds{B}^\mathsf{N} \rightarrow \mathds{B}^\mathsf{N}$, and
$\mathcal{P} \subset \mathds{N}^\ast$ a non empty and finite set of integers.
Any couple $(u,v) \in \mathcal{S}_{\llbracket 1, \mathsf{N} \rrbracket} \times \mathcal{S}_\mathcal{P}$ defines
a ``chaotic iterations based'' PRNG, which is denoted by  $\textit{CIPRNG}_f^2(u,v)$~\cite{wbg10:ip}. It is 
defined as follows:
\begin{equation}
\label{CIPRNGver2}
\left\{
\begin{array}{l}
 x^0 \in \mathds{B}^\mathsf{N}\\
 \forall n \in \mathds{N}, \forall i \in \llbracket 1, \mathsf{N} \rrbracket, x_i^{n+1} = \left\{ \begin{array}{ll} f(x^n)_i & \text{if  }~ i=u^n \\ x_i^n & \text{else} \end{array} \right.\\
 \forall n \in \mathds{N}, y^n = x^{v^n} .
\end{array}
\right.
\end{equation}
The outputted sequence produced by this generator is $\left(y^n\right)_{n \in \mathds{N}}$. 

The formerly proposed $\textit{CIPRNG}_f^1(u)$~\cite{bgw09:ip,guyeuxTaiwan10} is equal to \linebreak $\textit{CIPRNG}_f^2\left(u,\left(1\right)_{n\in \mathds{N}}\right)$, where $\left(1\right)_{n\in \mathds{N}}$ is the sequence that is  uniformly equal to 1. 
It has been proven as chaotic for the vectorial Boolean negation $f_0:\mathds{B}^\mathsf{N} \longrightarrow \mathds{B}^\mathsf{N}$, 
$(x_1, \hdots , x_\mathsf{N}) \longmapsto (\overline{x_1}, \hdots, \overline{x_\mathsf{N}})$ in \cite{bgw09:ip} 
and for a larger set of well-chosen iteration functions in~\cite{bcgr11:ip} but,
as only one bit is modified at each iteration, this generator is not able to pass any reasonable statistical tests.
The $\textit{XOR~CIPRNG}(S)$, for its part~\cite{DBLP:journals/corr/abs-1112-5239}, is defined as follows: $x^0 \in \mathds{B}^\mathsf{N}$, and $\forall n \in \mathds{N}, x^{n+1} = x^n \oplus S^n$
where $S \in \mathcal{S}_{\llbracket 1, \mathsf{N} \rrbracket}$ and $\oplus$ stands for the bitwise \emph{exclusive or} (xor) operation
between the binary decomposition of $x^n$ and $S^n$. This is indeed a $CIPRNG_{f_0}^2 (u,v)$ generator:
for any given $S \in \mathcal{S}_{\llbracket 1, \mathsf{N} \rrbracket}$, $v^n$ is the number
of 1's in the binary decomposition of $S^n$ while $u^{v^n}, u^{v^n+1}, \hdots , u^{v^{n+1}-1}$
are the positions of these ones.
The $\textit{XOR~CIPRNG}$ has been proven chaotic and it is able to pass all the most stringent statistical 
batteries of tests~\cite{DBLP:journals/corr/abs-1112-5239}, namely the well-known DieHARD, NIST, and TestU01. Furthermore, the output sequence is cryptographically secure
when $S$ is cryptographically secure~\cite{DBLP:journals/corr/abs-1112-5239}.
Following the same canvas than in the previous section, we have then characterized which $\textit{CIPRNG}_f^2(u,v)$ is chaotic according to Devaney.

%

Denote by $\mathcal{X}_{\mathsf{N},\mathcal{P}}=  \mathds{B}^\mathsf{N} \times \mathds{S}_{\mathsf{N},\mathcal{P}}$, where 
$\mathds{S}_{\mathsf{N},\mathcal{P}}=\mathcal{S}_{\llbracket 1, \mathsf{N} \rrbracket}\times \mathcal{S}_\mathcal{P}$.  
We then introduce a directed graph $\mathcal{G}_{f,\mathcal{P}}$ as follows.
\begin{itemize}
\item Its vertices are the $2^\mathsf{N}$ elements of $\mathds{B}^\mathsf{N}$.
\item Each vertex has $\displaystyle{\sum_{i=1}^\mathsf{p} \mathsf{N}^{p_i}}$ arrows, namely all the $p_1, p_2, \hdots, p_\mathsf{p}$ tuples 
having their elements in $\llbracket 1, \mathsf{N} \rrbracket $.
\item There is an arc labelled $a_1, \hdots, a_{p_i}$, $i \in \llbracket 1, \mathsf{p} \rrbracket$ between vertices $x$ and $y$ if and only if $y=F_{f,p_i} (x, (a_1, \hdots, a_{p_i})) $.
\end{itemize}
We have finally proven that 
\begin{theorem}
The pseudorandom number generator $\textit{CIPRNG}_f^2$ 
is chaotic on $\mathcal{X}_{\mathsf{N},\mathcal{P}}$ 
if
and only if its graph $\mathcal{G}_{f,\mathcal{P}}$ is strongly connected.
\end{theorem}

\section{Application to hash functions}
\label{Application to hash functions}

For the interest to add chaos properties to an hash function,
among other things regarding their diffusion and confusion, 
reader is referred to~\cite{bcg12:ij}.
%
%
%
Recall that, among other cryptographical properties, an hash function must be resistant to collisions: an adversary should not be able to find two distinct messages $m$ and $m'$ such that $h(m) = h(m')$.
%
%
%
%
Furthermore, an hash function must be second-preimage resistant, that is to say: an adversary given a message $m$ should not be able to find another message $m'$ such that $m\neq m'$ and $h(m)=h(m')$.

Let us now give a post-operative mode that can be applied to a cryptographically secure hash function without loosing 
the cryptographic properties recalled above.


\begin{definition}
Let 
\begin{itemize}
\item $k_1,k_2,n \in \mathds{N}^*$,
\item $h:(k,m) \in \mathds{B}^{k_1}\times\mathds{B}^* \longmapsto h(k,m) \in \mathds{B}^n$ a keyed hash function,
\item $S:k\in \mathds{B}^{k_2} \longmapsto \left(S(k)^i\right)_{i\in \mathds{N}} \in \llbracket 1,n\rrbracket^\mathds{N}$:
\begin{itemize}
\item either a cryptographically secure pseudorandom number generator (PRNG), 
\item or, in case of a binary input stream $m = m^0 || m^1 ||  m^1 || \hdots$ 
 where $\forall i, |m^i| = n$, $\left(S(k)^i\right)_{i\in \mathds{N}} = \left(m^k\right)_{i\in \mathds{N}}$.
%
\end{itemize}
\item $\mathcal{K}=\mathds{B}^{k_1}\times\mathds{B}^{k_2}\times \mathds{N}$ called the \emph{key space}, 
\item and $f:\mathds{B}^n \longrightarrow \mathds{B}^n$ a bijective map.
\end{itemize}
We define the keyed hash function $\mathcal{H}_h:\mathcal{K}\times\mathds{B}^* \longrightarrow \mathds{B}^n$ by the following procedure\\
\begin{tabular}{ll}
\underline{\textbf{Inputs:}} & $k = (k_1,k_2,n)\in \mathcal{K}$\\
                            & $m \in \mathds{B}^*$\\
\underline{\textbf{Runs:}} & $X=h(k_1,m)$, or $X=h(k_1,m^0)$ if $m$ is a stream\\
                           & for $i=1, \hdots, n:$\\
                           & ~~~~ $X=G_f(S^i,X)$\\
                           & return $X$
\end{tabular}
\end{definition}

$\mathcal{H}_h$ is thus a chaotic iteration based post-treatment 
on the inputted hash function $h$.
The strategy is provided by a secured PRNG when the machine operates
in a vacuum whereas it is redetermined at each iteration from the
input stream in case of a finite machine open to the outside world.
By doing so, we obtain a new hash function $\mathcal{H}_h$ with $h$, 
and this new one has a chaotic dependence regarding the inputted stream.
Furthermore, we have stated that~\cite{gwfb14:oip},
\begin{theorem}
If $h$ satisfies the collision resistance property, then it is the case too for $\mathcal{H}_h$.
And if $h$ satisfies the second-preimage resistance property, then it is the case too for $\mathcal{H}_h$.
\end{theorem}

Finally, as $\mathcal{H}_h$ simply operates chaotic iterations with strategy $\mathcal{S}$ provided at each iterate by the media, we have~\cite{gwfb14:oip}:
\begin{theorem}
In case where the strategy $\mathcal{S}$ is the bitwise XOR between a secured PRNG and the input stream,
the resulted hash function  $\mathcal{H}_h$ is chaotic.
\end{theorem}

\section{Application to symmetric cryptography}
\label{Application to symmetric cryptography}

Let us now present our last discoveries in the field of chaotic iteration based security. We have recently proven that the well-known Cipher Block Chaining (CBC) mode of operation, invented by IBM in 1976, can behave chaotically. The demonstration of this result is outlined thereafter, while details regarding the CBC mode of operation can be found in Patent~\cite{ehrsam1978message}.

Let us consider $\mathcal{X}=\mathds{B}^\mathsf{N}\times\mathcal{S}_\mathsf{N}$, where:
\begin{itemize}
\item $\mathsf{N}$ is the size for the block cipher,
\item $\mathcal{S}_\mathsf{N} = \llbracket 0, 2^\mathsf{N}-1\rrbracket^\mathds{N}$, the set of infinite sequences of natural integers bounded by $2^\mathsf{N}-1$, or the set of infinite $\mathsf{N}$-bits block messages,
\end{itemize}
in such a way that $\mathcal{X}$ is constituted by couples of internal states of the mode of operation together with sequences of block messages.
Let us consider the initial function:
$$\begin{array}{cccc}
 i:& \mathcal{S}_\mathsf{N} & \longrightarrow & \llbracket 0, 2^\mathsf{N}-1 \rrbracket \\
 & (m^i)_{i \in \mathds{N}} & \longmapsto & m^0
\end{array}$$
that returns the first block of a (infinite) message, and the shift function:
$$\begin{array}{cccc}
 \sigma:& \mathcal{S}_\mathsf{N} & \longrightarrow & \mathcal{S}_\mathsf{N} \\
 & (m^0, m^1, m^2, ...) & \longmapsto & (m^1, m^2, m^3, ...)
\end{array}$$
which removes the first block of a message. Let $m_j$ be the $j$-th bit of integer, or block message, $m\in \llbracket 0, 2^\mathsf{N}-1 \rrbracket$ expressed in the binary numeral system, and when counting from the left. We define:
$$\begin{array}{cccc}
F_f:& \mathds{B}^\mathsf{N}\times \llbracket 0, 2^\mathsf{N}-1 \rrbracket & \longrightarrow & \mathds{B}^\mathsf{N}\\
 & (x,m) & \longmapsto & \left(x_j m_j + f(x)_j \overline{m_j}\right)_{j=1..\mathsf{N}}
\end{array}$$
This function returns the inputted binary vector $x$, whose $m_j$-th components $x_{m_j}$ have been replaced by $f(x)_{m_j}$, for all $j=1..\mathsf{N}$ such that $m_j=1$. 
So the CBC mode of operation can be rewritten as the following dynamical system:
\begin{equation}
\label{eq:sysdyn}
\left\{
\begin{array}{ll}
X^0 = & (IV,m)\\
X^{n+1} = & \left(\mathcal{E}_k \circ F_{f_0} \left( i(X_1^n), X_2^n\right), \sigma (X_1^n)\right)
\end{array}
\right.
\end{equation}
where $IV$ is the input vector, $m$ the message to encrypt, and $\mathcal{E}_k$ the keyed symmetric cypher which has been selected. For $g:\llbracket 0, 2^\mathsf{N}-1\rrbracket \times \mathds{B}^\mathsf{N} \longrightarrow \mathds{B}^\mathsf{N}$, we denote $G_g(X) = \left(g(i(X_1),X_2);\sigma (X_1)\right)$. So the reccurent relation of Eq.\eqref{eq:sysdyn} can be rewritten in a condensed way, as follows.
\begin{equation}
X^{n+1} = G_{\mathcal{E}_k\circ F_{f_0}} \left(X^n\right) .
\end{equation}


Using all this material, we have proven that:
\begin{theorem}
\label{prop:transitivity}
Let $g=\varepsilon_k \circ F_{f_0}$, where $\varepsilon_k$ is a given keyed block cipher and $f_0:\mathds{B}^\mathsf{N} \longrightarrow \mathds{B}^\mathsf{N}$, $(x_1,...,x_\mathsf{N}) \longmapsto (\overline{x_1},...,\overline{x_\mathsf{N}})$ is the Boolean vectorial negation.
We consider the directed graph $\mathcal{G}_g$, where:
\begin{itemize}
\item vertices are all the $\mathsf{N}$-bit words.
\item there is an edge $m \in \llbracket 0, 2^{\mathsf{N}}-1 \rrbracket$ from $x$ to $\check{x}$ if and only if $g(m,x)=\check{x}$.
\end{itemize}
So if $\mathcal{G}_g$ is strongly connected, then $G_g$ is strongly transitive, and so the CBC mode of operation is chaotic.
\end{theorem}


\section{Conclusion}
\label{Conclusions}
In this article, the research works we have previously done in the field of chaotic iterations are summarized and clarified. Applications for pseudorandom number generation, hash function, and symmetric cryptography have then been outlined. Both theoretical analysis and experimental results confirm the feasibility of this approach.

\bibliographystyle{plain}
\bibliography{mabase}

\end{document}